\documentclass[aps,reprint,showpacs,floatfix,superscriptaddress,amsmath,amssymb,prb]{revtex4-1}
\usepackage{graphicx}
\usepackage{multirow}

\newcommand{\ket} [1] {| #1 \rangle}
\newcommand{\bra} [1] {\langle #1 |}
\newcommand{\braket}[2]{\langle #1 | #2 \rangle}
\newcommand{\ketbra}[2]{\ket{#1}\bra{#2}}
\newcommand{\lmax}{\lambda_{\tiny \mbox{max}}}
\newcommand{\specialcell}[2][c]{\begin{tabular}[#1]{@{}c@{}}#2\end{tabular}}
\newcommand{\smallmat}[4]{\bigl(\begin{smallmatrix}#1&#2\\#3&#4\end{smallmatrix}\bigr)}
\newcommand{\Psiproj}{\ket{\Psi^{\tiny \mbox{proj}}}}
\newcommand{\Aitriv}{\hat{A}_i^{\tiny \mbox{triv}}}
\newcommand{\Aiproj}{\hat{A}_i^{\tiny \mbox{proj}}}
\newcommand{\Atriv}{\hat{A}^{\tiny \mbox{triv}}}
\newcommand{\Aproj}{\hat{A}^{\tiny \mbox{proj}}}
\newcommand{\Vg}[1]{\hat{V}_g^{#1}}
\newcommand{\Aw}[1]{\hat{A}^{#1}}
\newcommand{\Aiw}[1]{\hat{A}_i^{#1}}

\newcommand{\eref}[1]{Eq.~(\ref{#1})}
\newcommand{\fref}[1]{Fig.~\ref{#1}}

\usepackage{array}

\begin{document}
\title{Identifying quantum phases from injectivity of symmetric matrix product states}
\author{Sukhwinder Singh}
\affiliation{Center for Engineered Quantum Systems, Department of Physics \& Astronomy, \\Macquarie University, 2109 NSW, Australia}

\begin{abstract}
Given a local gapped Hamiltonian with a global symmetry on a one dimensional lattice we describe a method to identify if the Hamiltonian belongs to a quantum phase in which the symmetry is spontaneously broken in the ground states or to a specific symmetry protected phase, without using local or string order parameters. We obtain different matrix product state (MPS) descriptions of the \textit{symmetric} ground state(s) of the Hamiltonian by \textit{restricting} the MPS matrices to transform under different equivalence classes of projective representations of the symmetry. The phase of the Hamiltonian is identified by examining which MPS descriptions, if any, are \textit{injective}, namely, whether the largest eigenvalue of the \textit{transfer matrix} obtained from the MPS is unique. We demonstrate the method for translation invariant Hamiltonians with a global SO(3), $Z_2$ and $Z_2 \times Z_2$ symmetry on an infinite chain.
\end{abstract}
 

\maketitle
Quantum many-body systems exhibit a variety of phases at zero temperature, and identifying the quantum phases that appear in a given system---to determine the phase diagram of the system---is a pivotal task e.g., in condensed matter physics.
In the absence of symmetries, all local gapped \cite{localgap} Hamiltonians on a one dimensional (1D) lattice belong to the same phase, and can be smoothly connected \cite{smoothness} to a ``trivial'' Hamiltonian whose ground state is a product state. In two or higher dimensions \textit{topological phases}, characterized by ground states with non-zero topological entanglement entropy\cite{topological}, can also appear even in the absence of symmetries. But such phases do not exist in 1D systems \cite{Wen2011, Schuch2011}.
A 1D local gapped Hamiltonian with a global symmetry can belong either to a \textit{symmetry broken} phase, characterized by degenerate ground states that are not all symmetric, or to one of possibly several distinct \textit{symmetry protected} phases, in which the ground state is unique and symmetric \cite{Pollmann2010,Wen2011,Schuch2011}.


Symmetry breaking can be identified using a local order parameter \cite{LOP} while string (non-local) order parameters have been proposed \cite{Nijs89,Haegeman12,Duivenvoorden12} to distinguish certain symmetry protected phases. A classical simulation of the system can select the symmetric ground state in a symmetry broken phase, or artificially break the symmetry in a symmetry protected phase due to numerical errors; in these cases the phase can no longer be identified by the corresponding local or string order parameter respectively. In this paper we introduce a method to identify quantum phases in classical simulations of 1D quantum many-body systems, without using local or string order parameters.

Ground states of 1D local gapped Hamiltonians can be efficiently described as \textit{matrix product states}\cite{Fannes92,Verstraete06,Hastings07,Garcia07} (MPSs). Quantum phases in 1D have also been classified \cite{Wen2011,Schuch2011} using MPS description of ground states, which has led to practical procedures for identifying \cite{Haegeman12,Pollmann2012, Li13} phases in classical simulations. Here we describe how to identify both symmetry broken and symmetry protected phases by examining the degeneracy of the largest eigenvalue of the transfer matrix obtained from MPS descriptions of \textit{symmetric} ground states. The latter are obtained by explicitly \textit{restricting} the simulation to the symmetric subspace of the lattice \cite{McCulloch02, Singh100, Singh101, Weichselbaum12} (even when the symmetry is spontaneously broken in the ground states).

\begin{figure}
  \includegraphics[width=8cm]{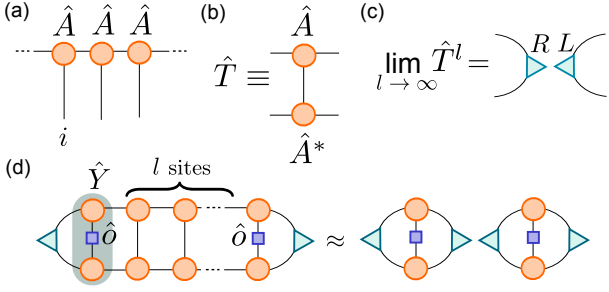}
\caption{\label{fig:1} (Color online) (a) Translation invariant MPS $\{\hat{A}_i\}$ on an infinite lattice, \eref{eq:1}. (b) Transfer matrix $\hat{T}$, \eref{eq:3}. (c) Condition satisfied by an injective MPS, $\lim_{l \to \infty} \hat{T}^l = \ket{R}\bra{L}$, where $\ket{R},\bra{L}$ are the dominant right and left eigenvectors of $\hat{T}$. (d) For an injective MPS we have $\langle\hat{o}_m \hat{o}_n\rangle \approx \langle\hat{o}_m\rangle \langle\hat{o}_n\rangle$ for sufficiently large $|m-n|=l$ [\eref{eq:2}].}
\end{figure}

Consider a 1D lattice $\mathcal{L}$ made of $L$ sites each described by a $d$-dimensional vector space $\mathbb{V}$. An translation invariant matrix product state $\ket{\Psi}$ of $\mathcal{L}$ can be expanded as [\fref{fig:1}(a)]
\begin{equation}\label{eq:1}
\ket{\Psi} = \sum_{i_1\ldots i_L=1}^{d} \mbox{Tr}(\hat{A}_{i_1}\hat{A}_{i_2}\ldots\hat{A}_{i_L}) \ket{i_1,i_2,\ldots,i_L},
\end{equation}
where Tr denotes matrix trace, $\ket{i_1,i_2,\ldots,i_L} = \bigotimes_{k \in \mathcal{L}}\ket{i_k}$, $\ket{i_k}$ is a local basis on site $k$, and $\hat{A}_{i_k}$ are site-independent, $\chi\times\chi$ matrices acting on vector space $\mathbb{W}$, $\hat{A}_{i_k} : \mathbb{W} \rightarrow \mathbb{W}$. Here $\chi$ is called the \textit{bond dimension} of the MPS.
We will assume that the MPS is in the \textit{canonical form} in which matrices $\hat{A}_i$ satisfy $\sum_i\hat{A}_i\hat{A}^\dagger_i =\hat{I}$ \cite{Garcia07,cfunique}. In this paper we consider the thermodynamic limit, $L\rightarrow \infty$, in order to accommodate symmetry breaking.

In a translation invariant MPS $\hat{A}$, two point correlations, $C(l)\equiv \langle\hat{o}_m \hat{o}_n\rangle - \langle\hat{o}_m\rangle \langle\hat{o}_n\rangle$, can be obtained as
\begin{equation}\label{eq:2}
C(l) = \mbox{Tr}(\hat{Y}\hat{T}^{l}\hat{Y}\hat{T}^{L-l-2})-\mbox{Tr}(\hat{Y}\hat{T}^{L-1})\mbox{Tr}(\hat{Y}\hat{T}^{L-1}),
\end{equation}
where $\hat{Y} \equiv \sum_{ij} \hat{o}_{ij} A_{i} \otimes A^*_{j}$, $|m-n|=l+1$, and
\begin{equation}\label{eq:3}
\hat{T} \equiv \sum_{i=1}^d A_{i} \otimes A^*_{i}
\end{equation}
is the \textit{transfer matrix}, see \fref{fig:1}.

If the largest modulus eigenvalue $\lmax$ of $\hat{T}$ is unique then the MPS is said to be \textit{injective} \cite{injective}.
For an injective MPS $\lmax=1$ \cite{Garcia07} and $\lim_{l \to \infty} \hat{T}^l =$ $\ket{R}\bra{L}$ where $\ket{R},\bra{L}$ are the right and left eigenvectors of $\hat{T}$ corresponding to $\lmax$ respectively [\fref{fig:1}(c)]. 
For sufficiently large but finite $l$ we have $\hat{T}^{l} \approx \ket{R}\bra{L}$ up to $O(|\lambda_2|^l)$ corrections where $\lambda_2$ is the second largest eigenvalue of $\hat{T}$. This implies that an injective MPS has a finite correlation length $\xi=-\frac{1}{\mbox{ln}|\lambda_2|}$ since $C(l) \approx e^{-l/\xi}$ [\fref{fig:1}(d)].
On the other hand, a \textit{non-injective} MPS (where $\lmax$ is degenerate) can have long-range correlations. For example, the MPS comprised of matrices $\{\hat{A}^{\tiny \mbox{GHZ}}_i, i=1,2,\ldots d\}$, where $\hat{A}^{\tiny \mbox{GHZ}}_i$ is a $d \times d$ matrix with 1 at position $(i,i)$ and 0 elsewhere, is non-injective and has long-range correlations. Specifically, it describes the GHZ state: $\frac{1}{\sqrt{d}}\sum_{i=1}^d \ket{i,i,\ldots,i}$, $\braket{i}{i'}=\delta_{ii'}$.

A state $\ket{\Psi}$ described by an injective MPS $\hat{A}$ can always be described by a non-injective MPS composed of matrices $\hat{A}'_i = \hat{I}_f \otimes \hat{A}_i$, where $\hat{I}_f$ is an $f \times f$ identity matrix. It is readily checked that correlations, \eref{eq:2}, of local observables obtained from the MPS $\hat{A}$ and $\hat{A}'$ are equal, up to a normalization factor $f$. The transfer matrix of MPS $\hat{A}'$ has $f^2$ eigenvalues with modulus 1, but describes a state with a finite correlation length. We will say that $\hat{A}'$ is an \textit{inflated} MPS description of state $\ket{\Psi}$.


Let us introduce the action of a symmetry group $\mathcal{G}$ on the lattice $\mathcal{L}$ by means of a unitary linear representation $\hat{U}_g : \mathbb{V} \rightarrow \mathbb{V}$ on each site $\mathbb{V}$, $\hat{U}_g\hat{U}_h = \hat{U}_{g.h}$ ($\forall g,h \in \mathcal{G}$). MPS $\ket{\Psi}$, \eref{eq:1}, has a global symmetry $\mathcal{G}$, or equivalently $\ket{\Psi}$ is $\mathcal{G}$-symmetric, if
\begin{equation}\label{eq:4}
\ket{\Psi} = (\bigotimes_{s \in \mathcal{L}} \hat{U}_{g}) \ket{\Psi}, ~~~ \forall g \in \mathcal{G}.
\end{equation}
The global symmetry implies a constraint on matrices $\hat{A}_i$, namely, $\ket{\Psi}$ is $\mathcal{G}$-symmetric iff matrices $\hat{A}_i$ satisfy \cite{Garcia08, Sanz09}
\begin{equation}\label{eq:5}
\sum_{i'}(\hat{U}_g)_{ii'}\hat{A}_{i'} = e^{i\theta_g}\hat{V}_g^{\dagger} \hat{A}_i \hat{V}_g, ~~~\forall g \in \mathcal{G},
\end{equation}
where the phases\cite{u1phases} $e^{i\theta_g}$ form a one dimensional representation of $\mathcal{G}$ and $\hat{V}_g:\mathbb{W}\rightarrow\mathbb{W}$ are unitary matrices (for an MPS in the canonical form) that form a $\chi$-dimensional \textit{projective} representation of $\mathcal{G}$---a representation that fulfills the group product only up to a phase, $\hat{V}_{g}\hat{V}_{h} = e^{i\omega(g,h)}\hat{V}_{g.h}$ $\forall g,h \in \mathcal{G}$, see App.\ref{sec:A1}. We refer to $\hat{V}_g$ as the \textit{bond representation} of the $\mathcal{G}$-symmetric MPS $\hat{A}$.

We now turn to addressing the goal of this paper. We have a local, gapped, translation invariant \cite{TIHAM} and $\mathcal{G}$-symmetric Hamiltonian $\hat{H}$ on the lattice $\mathcal{L}$ i.e.,
\begin{equation}\label{eq:6}
[\hat{H},\bigotimes_{s \in \mathcal{L}} \hat{U}_{g}] = 0,~\forall g \in \mathcal{G}.
\end{equation}
Our goal is to identify if $\hat{H}$ belongs to one of possibly several phases protected by symmetry $\mathcal{G}$ or to a phase in which symmetry $\mathcal{G}$ is broken in the ground states.

\textbf{Identification of symmetry protected phases.} If the ground state of $\hat{H}$ is unique and $\mathcal{G}$-symmetric then $\hat{H}$ belongs to a quantum phase protected by the symmetry $\mathcal{G}$. Distinct symmetry protected phases are in one to one correspondence with the elements of the second cohomology group of $\mathcal{G}$, $H^2(\mathcal{G},U(1))$, which also label different equivalence classes of projective representations of $\mathcal{G}$. Linear representations of $\mathcal{G}$ [the identity element of $H^2(\mathcal{G},U(1))$] correspond to the trivial symmetry protected phase. For example, the second cohomology group of $\mathcal{G}=$ SO(3) is $Z_2$. Thus, there are 2 distinct phases protected by SO(3) symmetry: the trivial phase corresponding to integer spin representations (linear), and a phase corresponding to half-integer spin (projective) representations of SO(3). A ground state belonging to a symmetry protected phase has a finite correlation length \cite{Hastings07,Garcia07} and admits an injective MPS description \cite{InflateNotUsual}. If $\hat{H}$ belongs to a symmetry protected phase $\omega \in H^2(\mathcal{G},U(1))$ then an \textit{injective} MPS description of its ground state has a bond representation in the equivalence class $\omega$ \cite{Wen2011,Schuch2011}.

\begin{figure}
  \includegraphics[width=6.5cm]{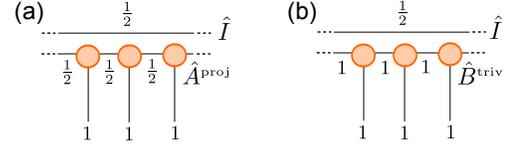}
\caption{\label{fig:2} (Color online) Illustration of an inflated MPS description in the (a) non-trivial and the (b) trivial SO(3) protected phase realized on a spin 1 chain. MPS $\Aproj$ and $\hat{B}^{\tiny \mbox{triv}}$ are SO(3)-symmetric and injective. Each index carries an irreducible spin representation as indicated. Identity $\hat{I}$ is depicted by a horizontal line (tensor product with the MPS). Change of basis $\hat{W}$ (see text) is eliminated for simplicity.}
\end{figure}

Consider a spin 1 Hamiltonian in the non-trivial SO(3) protected phase whose ground state $\Psiproj$ is described by an injective MPS $\Aproj$ with spin $\frac{1}{2}$ bond representation i.e., $\hat{V}_g$ in \eref{eq:5} is generated by the Pauli matrices [\fref{fig:2}(a)].
A simple example of such a ground state is the AKLT state \cite{AKLT} in the Haldane phase \cite{Haldane83}. State $\Psiproj$ can also be described by an \textit{inflated} MPS, for instance, comprised of matrices $\Aitriv \equiv \hat{W}^\dagger(\hat{I}_2\otimes \Aiproj)\hat{W}$ where $\hat{I}_2$ is the identity in the spin $\frac{1}{2}$ representation and $\hat{W}$ is the change of basis \cite{cfunique} from the tensor product of two spin $\frac{1}{2}$ representations, span$\{\ket{\uparrow\uparrow},\ket{\uparrow\downarrow},\ket{\downarrow\uparrow},\ket{\downarrow\downarrow}\}$, to the direct sum of spin 0 and spin 1 representation i.e., $\frac{1}{\sqrt{2}}(\ket{\uparrow\downarrow}-\ket{\downarrow\uparrow})\oplus$   span$\{\ket{\uparrow\uparrow}, \frac{1}{\sqrt{2}}(\ket{\uparrow\downarrow}+\ket{\downarrow\uparrow}),\ket{\downarrow\downarrow}\}$.

Thus, $\Psiproj$, which belongs to the non-trivial SO(3) protected phase, is also described by MPS $\Atriv$, which has integer spins bond representations (direct sum of spin 0 and spin 1). Note that this does not contradict the MPS based classification of symmetry protected phases because MPS $\Atriv$ is non-injective (inflated). Analogously, an injective MPS description of a ground state in the trivial SO(3) protected phase [e.g., MPS $\hat{B}^{\tiny \mbox{triv}}$ depicted in \fref{fig:2}(b)] has integer spins bond representation while MPS descriptions of the state with half-integer spins bond representation are inflated.

More generally, if $\hat{H}$ belongs to a symmetry protected phase $\omega \in H^2(\mathcal{G},U(1))$ then its ground state $\ket{\Psi}$ can be described by an MPS with a bond representation in any $\omega'$ different from $\omega$, but such an MPS description must be \textit{inflated} (see App. \ref{sec:A2}). More technically, an MPS description of $\ket{\Psi}$ with bond representation $\Vg{\omega'}$ in $\omega'$ is comprised of matrices $\Aiw{\omega'}\equiv\hat{W}^\dagger(\hat{I}_f \otimes \Aiw{\omega})\hat{W}$, 
where (i) $\Aw{\omega}$ is an injective MPS description of $\ket{\Psi}$ with bond representation $\Vg{\omega}$, (ii) $\hat{I}_f$ is the identity in a representation $\Vg{\tilde{\omega}}$, and (iii) $\hat{W}$ is the change of basis from the tensor product of representations $\Vg{\omega}$ and $\Vg{\tilde{\omega}}$ to the representation $\Vg{\omega'}$. ($\tilde{\omega}$ is chosen such that $\omega$ and $\omega'$ are related in this way.)
Thus, if we could obtain an MPS description of $\ket{\Psi}$ that satisfies \eref{eq:5} for a \textit{given} equivalence class of bond representation then we could iterate through the different equivalence classes $\omega \in H^2(\mathcal{G},U(1))$ and identify the phase of $\hat{H}$ from the $\omega$ that results in an injective MPS description of $\ket{\Psi}$. In MPS simulations, this can be achieved by choosing an initial $\mathcal{G}$-symmetric state with a bond representation in the given $\omega$ and ensuring that the symmetry, \eref{eq:5}, is protected in the simulation at all times.

In practice, an MPS description of the ground state(s) of a given Hamiltonian can be obtained e.g., by means of the Density Matrix Renormalization Group \cite{White92} (DMRG) and the Time-Evolving Block Decimation \cite{Vidal03} (TEBD) algorithms. One way to ensure that the DMRG and TEBD simulations produce a symmetric ground state is to incorporate the (necessary and sufficient) symmetry constraint \eref{eq:5} in the MPS ansatz.
It is well understood \cite{McCulloch02, Singh100, Singh101, Weichselbaum12} how to do this when the bond representation $\hat{V}_g$ is a \textit{linear} representation. When \eref{eq:5} involves linear representations, the matrices $\hat{A}_i$ decompose in terms of the Clebsch-Gordan (CG) coefficients of the group $\mathcal{G}$, which depend on the choice of the bond representation, and coefficients $\vec{x}$ that are not fixed by the symmetry (Wigner-Eckart theorem). An initial $\mathcal{G}$-symmetric MPS with a specific bond representation is constructed from the corresponding CG coefficients and randomly chosen $\vec{x}$. The symmetry is protected in each iteration of the DMRG and TEBD algorithms by only updating the $\vec{x}$ part of the MPS. We refer to Refs. \onlinecite{McCulloch02, Singh100, Singh101, Weichselbaum12} for details.

When the bond representation is projective, \eref{eq:5} can be incorporated in the MPS in the same way by exploiting the fact that projective representations of $\mathcal{G}$ can be lifted to \textit{linear} representations of another group $R(\mathcal{G})$, called the \textit{representation group}\cite{Boyle78} of $\mathcal{G}$. For example, $R$(SO(3))$=$SU(2) i.e., integer (linear) and half-integer (projective) spin representations of SO(3) are linear representations of SU(2). The group $R(\mathcal{G})$ is a central extension of $\mathcal{G}$ and in many cases of interest is also a covering group of $\mathcal{G}$. When \eref{eq:5} involves projective representations, the Wigner-Eckart decomposition of $\hat{A}_i$ is comprised of Clebsch-Gordan coefficients of the group $R(\mathcal{G})$.

To demonstrate the method consider the spin 1 bilinear biquadratic Heisenberg model on an infinite chain $\mathcal{L}$
\begin{equation}\label{eq:7}
\hat{H}^{\tiny \mbox{BLBQ}} = \sum_{k \in \mathcal{L}} \mbox{cos}\,\theta \left(\vec{S}_{k}\vec{S}_{k+1}\right) + \mbox{sin}\,\theta \left(\vec{S}_k\vec{S}_{k+1}\right)^2,
\end{equation}
where $\vec{S} \equiv (\hat{S}^x,\hat{S}^y,\hat{S}^z)$ are spin 1 matrices. This model has a global SO(3) symmetry and exhibits \cite{BLBQ, Haegeman12, Li13} the two distinct SO(3)-symmetry protected phases: There is a phase transition at $\theta = -\pi/4$ from the trivial phase, $\theta < -\pi/4$, to the Haldane phase \cite{Haldane83} corresponding to half-integer spin representations of SO(3).

For given $\theta$, we used the SU(2)-symmetric TEBD algorithm \cite{Singh100} to obtain two MPS descriptions of the SO(3)-symmetric ground state by restricting the bond representation to integer and half-integer spin representations respectively. In the two cases we chose an initial SO(3)-symmetric MPS with integer and half-integer spins bond representations respectively, which resulted in restricting the bond representation to these equivalence classes at all times in the SO(3)-symmetric simulation. This is because each site $\mathbb{V}$ of the lattice transforms as an integer spin representation, and both integer or half-integer bond representations (on space $\mathbb{W}$) correspond to a non-vanishing intertwiner (Clebsch-Gordan coefficients) between the spaces $\mathbb{V}\otimes\mathbb{W}$ and $\mathbb{W}$.
\begin{figure}
  \includegraphics[width=\columnwidth]{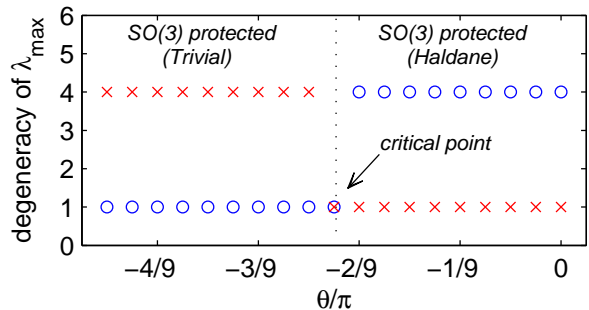}
\caption{\label{fig:3} (Color online) Degeneracy of $\lmax$ for two MPS descriptions of the SO(3)-symmetric ground states of $\hat{H}^{\tiny \mbox{BLBQ}}$, \eref{eq:7}, obtained by restricting the bond representation to integer ($\circ$) and half-integer ($\tiny \times$) spin representations respectively. Ground states were obtained using the SU(2)-symmetric  TEBD algorithm with $\chi \leq 100$.}
\end{figure}

From the plot in \fref{fig:3} we find that for $\theta < -\pi/4$ the MPS description of the ground state is injective for integer spins bond representation but inflated (with degeneracy of $\lmax$ equal to 4) for half-integer spins bond representation, and vice-versa for $\theta > -\pi/4$. Thus, we conclude that $\hat{H}^{\tiny \mbox{BLBQ}}$ belongs to the trivial phase for $\theta < -\pi/4$ and to the Haldane phase for $\theta > -\pi/4$. When the bond representation was restricted to integer spin representations in the Haldane phase [\fref{fig:3}] the simulation produced a \textit{minimally} inflated MPS, with a bond dimension that was $f=2$ times the injective MPS bond dimension, which corresponds to inflating the injective MPS by taking tensor product with identity in the spin $\frac{1}{2}$ representation. As a result, we find that the degeneracy of the largest eigenvalue of the transfer matrix of the inflated MPS is equal to $f^2=4$.

At the critical point $\theta = -\pi/4$ we find that the MPS description of the approximated ground state is injective when the bond representation is restricted to either integer or half-integer spin representations. At a critical point the ground state has a divergent correlation length which cannot be captured by an MPS with a finite bond dimension, and an MPS simulation only produces an approximation to the ground state---a ``nearby'' state lying in either gapped phase around the critical point. Here, in addition to a finite bond dimension, restricting the bond representation to integer or half-integer spin representations constrains the simulation to produce a nearby (injective) MPS lying in the Haldane or the trivial phase respectively.
\begin{figure}
  \includegraphics[width=\columnwidth]{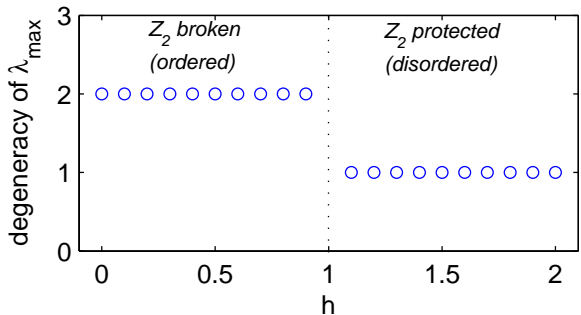}
\caption{\label{fig:4} (Color online) Degeneracy of $\lmax$ for the MPS description of $Z_2-$symmetric ground states of the Ising model, \eref{eq:8}. Ground states were obtained using the $Z_2-$symmetric TEBD algorithm with $\chi \leq 100$.}
\end{figure}

\textbf{Identification of symmetry broken phases.} If $\hat{H}$, \eref{eq:6}, belongs to a phase in which the global symmetry $\mathcal{G}$ is broken then it has a degenerate ground subspace and there exist ground states that are not $\mathcal{G}$-symmetric. More relevant to our purpose is that in a large class of symmetry broken phases there exist $\mathcal{G}$-\textit{symmetric} ground states all of which are GHZ-type states dressed with local entanglement (see App. \ref{sec:A3}), and  consequently their MPS descriptions are non-injective (for bond representations in any equivalence class).

For example, consider the spin-1/2 transverse field quantum Ising model on an infinite chain $\mathcal{L}$,
\begin{equation}\label{eq:8}
\hat{H}^{\tiny \mbox{ISING}} = \sum_{k \in \mathcal{L}} \hat{\sigma}_{k}^z \hat{\sigma}_{k+1}^z + h\hat{\sigma}_{k}^x,
\end{equation}
where $\hat{\sigma}^{z,x}$ are Pauli matrices and $h$ is the magnetic field in the transverse direction. This model has a $Z_2$ symmetry generated by a global spin flip, $\bigotimes_{k\in\mathcal{L}}\hat{\sigma}^x$. It exhibits a second-order phase transition at $h=1$ from the disordered phase---a trivial phase where the ground state is unique and $\mathbb{Z}_2$-symmetric---to the symmetry broken phase (ordered phase), $h<1$, where the ground state is 2-fold degenerate \cite{LOP}. For instance, at $h=0$ the ground subspace is spanned by states $\ket{\cdots\uparrow\uparrow\uparrow\cdots}$ and $\ket{\cdots\downarrow\downarrow\downarrow\cdots}$; there exist two $Z_2$-\textit{symmetric} ground states---$\frac{1}{\sqrt{2}}(\ket{\cdots\uparrow\uparrow\uparrow\cdots} \pm \ket{\cdots\downarrow\downarrow\downarrow\cdots})$--- which are GHZ states. In fact, $Z_2$-symmetric ground states throughout the symmetry broken phase contain GHZ-type correlations (App.\ref{sec:A3}), and consequently their MPS descriptions are non-injective. This is illustrated by the plot in \fref{fig:4}. We find that the MPS description of $Z_2$-symmetric ground states of the Ising model is non-injective (and non-inflated) for $h < 1$ and injective for $h > 1$, from which we infer that the symmetry is broken for $h < 1$.

\textbf{Example with $D_2 \cong Z_2 \times Z_2$ symmetry.} Finally, consider a lattice model that exhibits both a non-trivial symmetry protected phase and a symmetry broken phase. The spin 1 Heisenberg model on an infinite chain $\mathcal{L}$
\begin{equation}\label{eq:9}
\hat{H}^{\tiny \mbox{HEIS}} = \sum_{k \in \mathcal{L}} \vec{S}_{k}\vec{S}_{k+1} + D(S_{k}^z)^2
\end{equation}
has a global $D_2$ symmetry generated by rotations $\hat{R}^x = \mbox{exp}(i\pi\hat{S}^x)$ and $\hat{R}^z = \mbox{exp}(i\pi\hat{S}^z)$. Since $H^2(D_2,U(1))= Z_2$ there are 2 distinct $D_2$ symmetry protected phases, both exhibited by this model \cite{Gu09,Pollmann2012}. There is a phase transition at $D \approx 0.97$ \cite{Hu11} from the trivial (``large-D'') phase to the Haldane phase, $D > 0.97$, and another phase transition at $D \approx -0.3$ to an antiferromagnetic phase where the $D_2$ symmetry is broken to a $Z_2$ symmetry corresponding to the non-zero expectation value of $\hat{S}^z$.

\begin{figure}
  \includegraphics[width=\columnwidth]{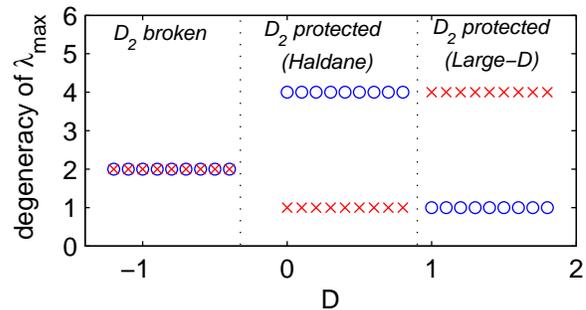}
\caption{\label{fig:5} (Color online) Degeneracy of $\lmax$ for two MPS descriptions of $D_2$-symmetric ground states of $\hat{H}^{\tiny \mbox{HEIS}}$, \eref{eq:9}, obtained by restricting the bond representation to linear ($\circ$) and projective ($\tiny \times$) representations of $D_2$ respectively. Ground states were obtained using the $R(D_2)$-symmetric TEBD algorithm (see App. \ref{sec:A4}) with $\chi \leq 100$. (The depicted phase boundaries are approximate.)}
\end{figure}

From the plot in \fref{fig:5} we find that in the large-$D$ phase the MPS description of the ground state is injective if the bond representation is linear but non-injective (inflated) if it is projective, and vice-versa in the Haldane phase. In the symmetry broken phase, MPS descriptions of the $D_2$-symmetric ground states are non-injective for linear or projective bond representation since both cases correspond to a GHZ-type state. 

\textbf{Outlook.}
The method presented here can be repeated in a symmetry broken phase to identify gapped phases that are protected by or break the residual symmetry by incorporating the residual symmetry in the MPS.
Symmetries are commonly incorporated in MPS algorithms to obtain computational speedup in simulations (see e.g., Refs.~\onlinecite{McCulloch02, Singh100, Singh101, Weichselbaum12}). The results presented here demonstrate that incorporating symmetries in MPS algorithms can also be useful to determine the gapped phase diagram of a 1D quantum many-body system.

%

\textit{Acknowledgements.-}
SS thanks Guifre Vidal, Frank Pollmann, and Gavin Brennen for inspiring and clarifying discussions, and also Mauro Cirio for useful conversations. SS acknowledges the hospitality of the Max-Planck Institute for Complex Systems and the Perimeter Institute for Theoretical Physics where this work was initiated.


%
%
\appendix
%
\section{Projective representations}
\label{sec:A1}
A (unitary) projective representation $\hat{V}_g$ of a group $\mathcal{G}$ fulfills the group product only up to a phase factor, $\hat{V}_{g}\hat{V}_{h} = e^{i\omega(g,h)}\hat{V}_{g.h},~\forall g,h \in \mathcal{G}$.

\textit{Example 1:} Consider the group $D_2 \cong Z_2 \times Z_2$ generated by rotations $\hat{R}^x = \mbox{exp}(i\pi\hat{S}^x)$ and $\hat{R}^z = \mbox{exp}(i\pi\hat{S}^z)$. The group product is
\begin{equation} 
g_{ij}. g_{mn} = g_{\tiny\mbox{mod}(i+m,2),\tiny\mbox{mod}(j+n,2)},~~~i,j,m,n\in\{0,1\}.\nonumber
\end{equation}
The representation $\hat{V}_{ij}$ of $D_2$ given by the Pauli matrices,
\begin{equation}
\hat{V}_{00}=\smallmat{1}{0}{0}{1},~\hat{V}_{01}=\smallmat{0}{1}{1}{0},~ \hat{V}_{10}=\smallmat{1}{0}{0}{-1},~\hat{V}_{11}=\smallmat{0}{-i}{i}{0},\nonumber
\end{equation}
is a projective representation since it fulfills the group product only up to a phase factor, \begin{equation}
\begin{split}
\hat{V}_{00}\hat{V}_{mn} &= \hat{V}_{mn},~\hat{V}_{01}\hat{V}_{10} = \hat{V}_{10}\hat{V}_{01}= i\hat{V}_{11},\\
\hat{V}_{01}\hat{V}_{11} &= i\hat{V}_{10},~\hat{V}_{11}\hat{V}_{01} = -i\hat{V}_{10},\nonumber
\end{split}
\end{equation}
which cannot be removed by scaling the representation matrices.

\textit{Example 2:} Half-integer spin representations are projective representations of SO(3). For example, in the spin $\frac{1}{2}$ representation, generated by $\hat{S}_i = \hat{\sigma_i}/2$ ($\sigma_i$ are the Pauli matrices), the composition of two $\pi$ rotations, say, around the $z$-axis is $e^{-i2\pi\hat{S}_z} = -\hat{I}$. Thus, the spin $\frac{1}{2}$ representation is a projective representation of SO(3) owing to the appearance of the factor $-1$.

\textit{Example 3:} The group $Z_n$ has no non-trivial projective representations.

A projective representation $\hat{V}_g$ of a group $\mathcal{G}$ is defined only up to a phase, $\hat{V}_g \leftrightarrow e^{i\phi_g} \hat{V}_g$, which results in equivalence classes of projective representations under the relation $\omega(g,h) \sim \omega(g,h) + \phi_g + \phi_h - \phi_{g.h}$ mod $2\pi$. The equivalence classes form a group that is isomorphic to the second cohomology group $H^2(\mathcal{G},U(1))$. A linear representation simply corresponds to $\omega(g,h) = 0$ for all $g,h$ in $\mathcal{G}$ and to the identity element of $H^2(\mathcal{G},U(1))$.

\section{Symmetric matrix product states in a symmetry protected phase \label{sec:A2}}

Consider a local, gapped and $\mathcal{G}$-symmetric Hamiltonian $\hat{H}$ on a one dimensional lattice that belongs to the symmetry protected phase corresponding to $\omega \in H^2(\mathcal{G},U(1))$. Any MPS description of the (unique) ground state $\ket{\Psi}$ of $\hat{H}$ is possibly (i) injective, (ii) GHZ-type non-injective, or (iii) inflated type non-injective. In this section we argue that an MPS description of $\ket{\Psi}$ that has a bond representation in an equivalence class $\omega' \in H^2(\mathcal{G},U(1)),~\omega' \neq \omega$, must be inflated [i.e. we will argue to rule out options (i) and (ii)]. This result was used in the paper to identify symmetry protected phases.

First, clearly $\ket{\Psi}$ cannot be described by a GHZ-type non-injective MPS with a bond representation in $\omega'$ since a GHZ-type non-injective MPS has long-range correlations while $\ket{\Psi}$ has short-range correlations.

Next, since $\ket{\Psi}$ is the unique ground state of a 1D local gapped Hamiltonian it can be described by an injective MPS \cite{Hastings07,Hastings071,Garcia07}. According to the MPS based characterization of symmetry protected phases, an injective MPS description of $\ket{\Psi}$ has a bond representation in the equivalence class $\omega \in H^2(\mathcal{G},U(1))$ \cite{Wen2011,Schuch2011}.

Let $\ket{\Psi}$ be described by an injective MPS $\hat{A}$. One may hope that the equivalence class ($\omega$) of the bond representation of MPS $\hat{A}$ may be changed by applying a unitary transformation $\hat{W}$ to the MPS matrices, $\hat{W}^\dagger\hat{A}_i\hat{W}$, thus defeating the MPS based characterization of symmetry protected phases. However, a simple argument shows that if MPS $\hat{A}$ and MPS $\hat{A}'_i = \hat{W}^\dagger \hat{A}_i \hat{W}$ describe the same $\mathcal{G}$-symmetric state then $\hat{W}$ must commute with $\mathcal{G}$,
\begin{equation}\label{eq:E}
\hat{V}_g \hat{W} = \hat{W} \hat{V}_g,~~~\forall g \in \mathcal{G}.
\end{equation}
[Consequently, $\hat{W}$ acts as a scalar matrix in the bond representation (Schur's lemma), and cannot e.g., map a projective representation in one equivalence class to a projective representation in another equivalence class.] This can be derived as follows. Matrices $\hat{A}'_i$ must also satisfy Eq.~(5) (main text),
\begin{equation}\label{eq:E0}
\sum_{i'}(\hat{U}_g)_{ii'} \hat{A}'_i = \hat{V}_g^{\dagger} \hat{A}'_i \hat{V}_g,~~~\forall g \in \mathcal{G}.
\end{equation}
Substituting $\hat{A}'_i = \hat{W}^\dagger \hat{A}_i \hat{W}$ in \eref{eq:E0},
\begin{equation}\label{eq:E1}
\hat{W}^\dagger[\sum_{i'}(\hat{U}_g)_{ii'} \hat{A}_i] \hat{W} = \hat{V}_g^{\dagger} \hat{W}^\dagger \hat{A}_i \hat{W} \hat{V}_g,~~~\forall g \in \mathcal{G}.
\end{equation}
By multiplying $\hat{W}^\dagger[.]\hat{W}$ on both sides of Eq.~(5) (main text) we obtain
\begin{equation}\label{eq:E2}
\hat{W}^\dagger[\sum_{i'}(\hat{U}_g)_{ii'} \hat{A}_i] \hat{W} = \hat{W}^\dagger \hat{V}_g^{\dagger} \hat{A}_i \hat{V}_g \hat{W},~~~\forall g \in \mathcal{G}.
\end{equation}
From \eref{eq:E1} and \eref{eq:E2} we obtain \eref{eq:E}.

Thus, an MPS description of $\ket{\Psi}$ with a bond representation in $\omega' \neq \omega$ cannot be injective or GHZ-type non-injective. The only option left to obtain an MPS description with a bond representation in $\omega'$ is to inflate an injective MPS description of $\ket{\Psi}$ as described in the paper.

\section{Symmetric matrix product states in a symmetry broken phase}\label{sec:A3}
Consider an infinite lattice $\mathcal{L}$ where each site transforms as a $d$-dimensional unitary representation $\hat{U}_g$ of a \textit{discrete} group $\mathcal{G}$. Also consider a local, gapped, translation invariant and $\mathcal{G}$-symmetric Hamiltonian $\hat{H}$ on the lattice that belongs to a quantum phase in which the symmetry $\mathcal{G}$ is spontaneously broken in the ground states. That is, $\hat{H}$ has a degenerate ground subspace and there exist ground states that are not $\mathcal{G}$-symmetric. In this section we argue that if $\mathcal{G}$ is Abelian, or if $\mathcal{G}$ is a non-Abelian symmetry that is broken in a given way (specified later) then the MPS descriptions of the $\mathcal{G}$-\textit{symmetric} ground states are non-injective. This result was used in the paper to identify symmetry breaking phases. In one dimension, continuous global symmetries cannot be spontaneously broken in local gapped Hamiltonians in accordance with the Mermin-Wagner theorem, so we do not consider this case here. Also see e.g., Refs.~\onlinecite{Schuch2011,Fannes92} for a related discussion.

\textit{Lemma 1.} Consider a translation invariant state $\ket{\Psi}$ of the lattice $\mathcal{L}$ that is described by an injective (canonical) MPS $\hat{A}$. Let $\lambda$ denote the largest modulus eigenvalue of the matrix
\begin{equation}\label{eq:lem1}
\hat{Y}_g \equiv \sum_{ij=1}^d (\hat{U}_g)_{ij} \hat{A}_{i} \otimes \hat{A}^*_{j}.
\end{equation}
Then $|\lambda| \leq 1$ for any $g \in \mathcal{G}$ with equality iff $\ket{\Psi}$ is $\mathcal{G}$-symmetric.

This result is proved in Ref.~\onlinecite{Garcia08} as Lemma 1. $\square$

\begin{figure}
  \includegraphics[width=\columnwidth]{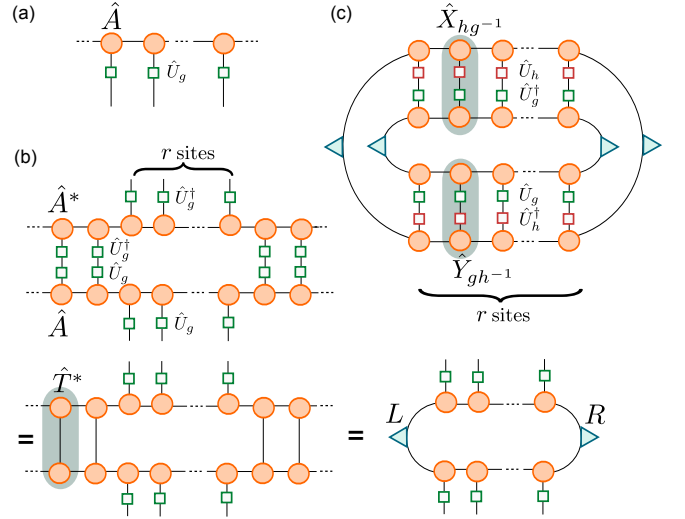}
\caption{\label{fig:app} (Color online) (a) State $\ket{\Psi_g}$, \eref{eq:lem2a}, as obtained by acting the symmetry on state $\ket{\Psi_e}$ described by an injective MPS $\hat{A}$. (b) Reduced density matrix $\hat{\rho_g}$ for $r$ sites in state $\ket{\Psi_g}$; $\ket{R},\bra{L}$ are the dominant right and left eigenvectors of the transfer matrix $\hat{T} = \sum_{i,j=1}^d A_{i} \otimes A^*_{j}$ respectively. Shown is the simplification of the expression for $\hat{\rho_g}$ by using $\hat{U}_g^\dagger\hat{U}_g = \hat{I}$ and $\lim_{l\to\infty}\hat{T}^l=\ketbra{R}{L}$. (c) $\mbox{Tr}(\hat{\rho}_g \hat{\rho}_h)$, \eref{eq:lem2b}.}
\end{figure}

\textit{Lemma 2.} Assume that there exists a ground state $\ket{\Psi_{e}}$ of $\hat{H}$ that is invariant only under the action of the identity element $e$ of $\mathcal{G}$ and that is described by an injective MPS $\hat{A}$. The state [\fref{fig:app}(a)]
\begin{equation}\label{eq:lem2a}
\ket{\Psi_g} \equiv (\bigotimes_{k \in \mathcal{L}} \hat{U}_g) \ket{\Psi_{e}},~~~g \neq e,
\end{equation}
is also a ground state of $\hat{H}$ (since $\hat{H}$ is $\mathcal{G}$-symmetric). Denote by $\hat{\rho}_g$ and $\hat{\rho}_h$ the reduced density matrices for $r$ sites in the states $\ket{\Psi_g}$ and $\ket{\Psi_h}$ respectively ($g,h \in \mathcal{G},~g \neq h$) [\fref{fig:app}(b)]. Then for sufficiently large $r$, the overlap of $\hat{\rho}_g$ and $\hat{\rho}_h$, $\mbox{Tr}(\hat{\rho}_g \hat{\rho}_h)$, is exponentially small (i.e., loosely speaking, ground states $\ket{\Psi_g}$ and $\ket{\Psi_h}$ become ``locally'' orthogonal after blocking $r$ sites of $\mathcal{L}$).

\textit{Proof:} The overlap of $\hat{\rho}_g$ and $\hat{\rho}_h$ is [\fref{fig:app}(c)]
\begin{equation}\label{eq:lem2b}
\mbox{Tr}(\hat{\rho}_g \hat{\rho}_h) = \bra{L}^{\otimes 2} (\hat{X}_{hg^{-1}}^{r} \otimes \hat{Y}_{gh^{-1}}^r) \ket{R}^{\otimes 2},
\end{equation}
where $\ket{L},\bra{R}$ are the dominant left and right eigenvectors of the transfer matrix $\hat{T} = \sum_{i,j=1}^d A_{i} \otimes A^*_{j}$ respectively, $\hat{X}_g \equiv \sum_{i,j=1}^d (\hat{U}_g^\dagger)_{ij} A_{i} \otimes A^*_{j}$, and $\hat{Y}_g$ is defined according to \eref{eq:lem1}.
Denote by $\lambda_{x}$ and $\lambda_y$ the largest modulus eigenvalue of matrices $\hat{X}_{hg^{-1}}$ and $\hat{Y}_{gh^{-1}}$ respectively. From lemma 1 it follows that $\lambda_x < 1,\lambda_y < 1$. This implies that for sufficiently large $r$ we have
\begin{equation}
\mbox{Tr}(\hat{\rho}_g \hat{\rho}_h) \approx O(\mbox{exp}({-\frac{r}{\xi_x}}) \mbox{exp}({-\frac{r}{\xi_y}})),
\end{equation}
where $\xi_x = -\frac{1}{\mbox{ln}\lambda_x}$ and $\xi_y = -\frac{1}{\mbox{ln}\lambda_y}$. $\square$

\textit{Lemma 3.} (\textit{Existence of $\mathcal{G}$-symmetric ground states.}) 
(a) If the group $\mathcal{G}$ is Abelian then $\hat{H}$ always has ground states that are $\mathcal{G}$-symmetric. (b) If $\mathcal{G}$ is non-Abelian $\hat{H}$ may not have any $\mathcal{G}$-symmetric ground states.

\textit{Proof (a).} Let lattice $\mathcal{L}$ be described by a (infinite dimensional) vector space $\mathbb{V}^{(\mathcal{L})}$. Under the action of the global symmetry $\mathcal{G}$, $\mathbb{V}^{(\mathcal{L})}$ decomposes as $\mathbb{V}^{(\mathcal{L})} \cong \bigoplus_{\alpha} \mathbb{V}_{\alpha}$ where $\alpha$ labels irreducible representations of $\mathcal{G}$. According to Schur's lemma the $\mathcal{G}$-symmetric Hamiltonian $\hat{H} : \mathbb{V}^{(\mathcal{L})} \rightarrow \mathbb{V}^{(\mathcal{L})}$ is block diagonal as
\begin{equation}\label{eq:lem3a}
\hat{H} = \bigoplus_{\alpha} \hat{H}_{\alpha},~~~\hat{H}_{\alpha} : \mathbb{V}_{\alpha} \rightarrow \mathbb{V}_{\alpha}.
\end{equation}
We can obtain eigenvectors of $\hat{H}$ in each symmetry sector $\alpha$ by diagonalizing each block $\hat{H}_{\alpha}$ separately. If $\mathcal{G}$ is Abelian then all irreps $\alpha$ are one dimensional. Clearly, all eigenvectors of $\hat{H}$ transform as a one dimensional irrep of $\mathcal{G}$ i.e., all eigenvectors are symmetric up to an overall phase. In particular, if the ground state is $n$-fold degenerate then there exist exactly $n$ $\mathcal{G}$-symmetric ground states $\{\ket{\Psi^{\tiny \mbox{sym}}_\alpha}\}$,
\begin{equation}\label{eq:lem3b}
\hat{U}_g\ket{\Psi^{\tiny \mbox{sym}}_\alpha}  = f_\alpha\ket{\Psi^{\tiny \mbox{sym}}_\alpha},~~~ \forall g \in \mathcal{G},f_\alpha \in \mathbb{C},~|f_\alpha|=1.
\end{equation}
If the symmetry is broken then there must exist at least two ground states that transform as \textit{different} one dimensional irreps of $\mathcal{G}$. This ensures that there exist superpositions of the two ground states, $a\ket{\Psi^{\tiny \mbox{sym}}_\alpha} + b\ket{\Psi^{\tiny \mbox{sym}}_{\alpha'}}$, that are non-symmetric since the two terms in the superposition pick up different phase factors $f_\alpha$ and $f_{\alpha'}$ under the action of the symmetry. $\square$

\textit{Proof (b).} If $\mathcal{G}$ is non-Abelian then the ground subspace can transform as an irrep $\alpha$ with dimension larger than one. In this case, and if no other ground states are present, clearly none of the ground states are $\mathcal{G}$-symmetric. $\square$

Finally, we argue that if the ground subspace of $\hat{H}$ is spanned by states that are mapped to one another by the action of the symmetry i.e., states $\{\ket{\Psi_g},g \in \mathcal{G}\}$ of lemma 2, then the $\mathcal{G}$-symmetric ground states $\{\ket{\Psi^{\tiny \mbox{sym}}}\}$ of $\hat{H}$ (lemma 3) are GHZ-type states, namely, equal probability superpositions of locally orthogonal states, generally after blocking the lattice. The latter implies that MPS descriptions of the $\mathcal{G}$-symmetric ground states of $\hat{H}$ are non-injective.

Let us block the lattice $\mathcal{L}$ such that states $\{\ket{\Psi_g}\}$, \eref{eq:lem2a}, become locally orthogonal (lemma 2). Since states $\{\ket{\Psi_g}\}$ span the ground subspace, a generic ground state $\ket{\Psi}$ of $\hat{H}$ can be expanded as
\begin{equation}\label{eq:proofa}
\ket{\Psi} = \sum_{h \in \mathcal{G}} c_{h} \ket{\Psi_h},~~~c_h \in \mathbb{C}.
\end{equation}
If state $\ket{\Psi}$ is $\mathcal{G}$-symmetric then $\ket{\Psi} = \hat{U}_g \ket{\Psi}~\forall g \in \mathcal{G}$, that is, 
\begin{equation}\label{eq:proofb}
\sum_{m \in \mathcal{G}} c_{m} \ket{\Psi_m} = \sum_{h \in \mathcal{G}} c_{h} \hat{U}_g\ket{\Psi_h}.
\end{equation}
Changing the dummy summation variable $m=g.h$ and using $\ket{\Psi_{g.h}} = \hat{U}_g\ket{\Psi_h}$ we obtain
\begin{equation}\label{eq:proofc}
\sum_{g.h \in \mathcal{G}} c_{g.h} \ket{\Psi_{g.h}} = \sum_{h \in \mathcal{G}} c_{h} \ket{\Psi_{g.h}}.
\end{equation}
It follows that $c_{g.h} = c_h, \forall g,h \in \mathcal{G}$ which implies $c_{g} = c_e, \forall g \in \mathcal{G}$. Thus, any $\mathcal{G}$-symmetric ground state $\ket{\Psi^{\tiny \mbox{sym}}}$ of $\hat{H}$ can be written as
\begin{equation}\label{eq:proofd}
\ket{\Psi^{\tiny \mbox{sym}}} = \sum_{g \in \mathcal{G}} c_e \ket{\Psi_g},
\end{equation}
where $c_e=\pm \frac{1}{\sqrt{|\mathcal{G}|}}$ (normalization). Thus, a $\mathcal{G}$-symmetric ground state $\ket{\Psi^{\tiny \mbox{sym}}}$ of $\hat{H}$ is a GHZ-type state (after blocking the lattice). $\square$

We interpret the plots in Fig. 4 and Fig. 5 (main text) to indeed indicate symmetry breaking resulting from the mechanism discussed above, namely, the symmetric ground states belonging to the symmetry broken phase exhibited in those models contain GHZ-type correlations and can be expanded according to \eref{eq:proofd}.

\section{$R(D_2)$-symmetric TEBD algorithm}\label{sec:A4}
The $D_2$-symmetric ground states used for the plot in Fig. 5 (main text) were obtained by means of the $R(D_2)$-symmetric version of the TEBD algorithm; $R(D_2)$ denotes the representation group \cite{Boyle78} of $D_2 = Z_2 \times Z_2$. The $R(D_2)$-symmetric TEBD algorithm was implemented by following Ref.~\onlinecite{Singh100} but replacing the irreps and Clebsch-Gordan coefficients of SU(2) with those of $R(D_2)$, which are summarized below.

$R(D_2)$ is a finite non-Abelian group. It has four one-dimensional irreps and one two-dimensional irrep, which we simply label as $\{0,1,2,3\}$ and $4$ respectively. The 1-d irreps correspond to linear irreps of $D_2$ and the 2-d irrep correponds to a projective representation of $D_2$ (see App.~\ref{sec:A1}). The Clebsch-Gordan (CG) rules for the direct sum decomposition of the tensor product of the various pairs of irreps of $R(D_2)$, symbolically
\begin{equation}
p\otimes q \cong \bigoplus r, ~~~p,q,r \in \{0,1,2,3,4\},\nonumber
\end{equation}
and the CG coefficients that describe the corresponding change of basis are summarized in Table \ref{table:1}.

\begin{table}[!t] \caption{Clebsch-Gordan coefficients for the group $R(D_2)$. $\{\hat{\sigma}_x, \hat{\sigma}_y,\hat{\sigma}_z,\hat{I}\}$: Pauli matrices; $\hat{I}_q$: identity in irrep $q$; $\gamma=\frac{1}{\sqrt{2}}$.}\label{table:1}
    \begin{tabular}{c c c}
    \hline\hline
    $\mathbf{p\otimes q}$ & $\mathbf{\bigoplus r}$ & \textbf{CG coeffs} \\ \hline
    $0 \otimes q$ & $q$ & $\hat{I}_q$ \\ \hline
    $q \otimes q,~q\neq 4$ & $0$ & 1 \\ \hline
    $4 \otimes 1$ & $4$ & $\hat{\sigma}_z$ \\ \hline
    $4 \otimes 2$ & $4$ & $\hat{\sigma}_y$ \\ \hline
    $4 \otimes 3$ & $4$ & $\hat{\sigma}_x$ \\ \hline
    $4 \otimes 4$ & $(0 \oplus 1 \oplus 2 \oplus 3)$ &  \specialcell{$0\rightarrow\small \gamma\hat{\sigma}_y$ \\
    $1\rightarrow \gamma\hat{\sigma}_x$ \\
    $2\rightarrow i\gamma\hat{I}$ \\
    $3\rightarrow -\gamma\hat{\sigma}_z$} \\ \hline
    \end{tabular}
\end{table}


\begin{thebibliography}{74}
\bibitem{localgap} \textit{Local} means that the many-body Hamiltonian is a sum of terms each of which only acts non-trivially on a small number of neighbouring sites, and \textit{gapped} means there is a finite energy difference between the ground subspace and the first excited state in the thermodynamic limit.

\bibitem{smoothness} By a \textit{smoothly connected} path of Hamiltonians we mean that ground state properties vary smoothly as the Hamiltonian is varied along the path.

\bibitem{topological} A. Kitaev and J. Preskill, Phys. Rev. Lett. \textbf{96}, 110404 (2006); M. Levin and X.-G. Wen,  Phys. Rev. Lett. \textbf{96}, 110405 (2006).
\bibitem{Wen2011} X. Chen, Z.-C. Gu, and X.-G. Wen, Phys. Rev. B \textbf{83}, 035107 (2011).
\bibitem{Schuch2011} N. Schuch, D. Perez-Garcia, I. Cirac, Phys. Rev. B \textbf{84}, 165139 (2011).
\bibitem{Pollmann2010}  F. Pollmann, A. M. Turner, Erez Berg, and Masaki Oshikawa, Phys. Rev. B \textbf{81}, 064439 (2010).

\bibitem{LOP} A local order parameter is the ground state expectation value of a local observable that does not commute with the symmetry. For example, in the 1D quantum Ising model---which has a global spin flip symmetry---the local order parameter is the ground state spin magnetization: it is zero in the disordered phase and non-zero in the ordered phase (which breaks the symmetry).

\bibitem{Nijs89} M. den Nijs and K, Rommelse, Phys. Rev. B \textbf{40}, 4709 (1989).
\bibitem{Haegeman12} J. Haegeman, D. P.-Garcia, I. Cirac and N. Schuch, Phys. Rev. Lett. \textbf{109}, 050402 (2012).
\bibitem{Duivenvoorden12} K. Duivenvoorden and T. Quella, Phys. Rev. B \textbf{86}, 235142 (2012); Phys. Rev. B \textbf{87}, 125145 (2013).
\bibitem{Fannes92} M. Fannes, B, Nachtergaele and R. Werner, Commun. Math. Phys. \textbf{144}, 443 (1992).
\bibitem{Verstraete06} F. Verstraete and J.I. Cirac, Phys. Rev. B \textbf{73}, 094423 (2006).
\bibitem{Hastings07} M. B. Hastings, J. Stat. Mech., P08024 (2007); Phys. Rev. B \textbf{76}, 035114 (2007).
\bibitem{Hastings071} M. B. Hastings, Phys. Rev. B \textbf{76}, 035114 (2007).
\bibitem{Garcia07} D. Perez-Garcia, F. Verstraete, M.M. Wolf, and J.I. Cirac, Quantum Inf. Comput. \textbf{7}, 401 (2007).
\bibitem{Pollmann2012} F. Pollmann and A. M. Turner, Phys. Rev. B \textbf{86}, 125441 (2012).
\bibitem{Li13} W. Li, A. Weichselbaum J. von Delft, Phys. Rev. B \textbf{88}, 245121 (2013).

\bibitem{McCulloch02} I. P. McCulloch and M. Gulacsi, Europhys. Lett. \textbf{57}, 852 (2002).
\bibitem{Singh100} S. Singh, H.-Q. Zhou, and G. Vidal, New J. Phys. \textbf{12}, 033029 (2010).
\bibitem{Singh101} S. Singh, R.N.C. Pfeifer and G. Vidal, Phys. Rev. A \textbf{82}, 050301 (2010); S. Singh, R.N.C. Pfeifer, G. Vidal and G. Brennen, Phys. Rev. B \textbf{89}, 075112 (2014).
\bibitem{Weichselbaum12} A. Weichselbaum, Annals of Physics \textbf{327} 2972-3047 (2012).

\bibitem{cfunique} For a given bond dimension, the canonical form of the MPS is unique up to unitary transformations $\hat{V} : \mathbb{W} \rightarrow \mathbb{W}$ i.e., state $\ket{\Psi}$ can be equivalently described by a canonical MPS comprised of matrices $\hat{V}^\dagger \hat{A}_i \hat{V}$. Other equivalent (canonical) MPS descriptions of $\ket{\Psi}$ are obtained by inflating the bond dimension as described in the main text.

\bibitem{injective} This means there exists $n \in \mathbb{Z}^+$ such that the map $\Gamma_n(\hat{X}) =$ $\sum_{i_1,\ldots,i_n} \mbox{Tr}(\hat{X}\hat{A}_{i_1}\cdots\hat{A}_{i_n})$ $\ket{i_1,\ldots,i_n}$ is injective, see also Ref.~\onlinecite{Garcia07}.

\bibitem{Garcia08} D. Perez-Garcia, M.M. Wolf, M. Sanz, F. Verstraete, and J.I. Cirac, Phys. Rev. Lett. \textbf{100}, 167202 (2008).
\bibitem{Sanz09} M. Sanz, M. M. Wolf, D. Pérez-García, and J. I. Cirac, Phys. Rev. A \textbf{79}, 042308 (2009).

\bibitem{u1phases} In certain cases the factor $e^{i\theta_g}$ that appears in \eref{eq:5} leads to a further classification of symmetry protected phases \cite{Wen2011}. In this paper we do not consider these cases and ignore $e^{i\theta_g}$ in \eref{eq:5}.


\bibitem{TIHAM} Here we consider translationally invariant Hamiltonians for simplicity, but we expect that our results also apply to non-translationally invariant systems. In the latter, the transfer matrix becomes site dependent and injectivity of MPS descriptions is diagnosed by examining the eigenvalues of the transfer matrix for each site of the lattice.

\bibitem{InflateNotUsual} The unique ground state of a local gapped Hamiltonian can also be described by an inflated (non-injective) MPS. However, MPS simulations do not usually produce inflated descriptions. This is also not desired in practice since an inflated MPS approximates the ground state with a lower accuracy as compared to an injective MPS with the same bond dimension. One can try to detect artificially inflated MPS descriptions by checking if the simulation continues to produce an inflated MPS after decreasing the bond dimension. On the other hand, in this paper we describe how MPS simulations can be constrained---by enforcing suitable symmetry constraints on the MPS---to produce inflated MPS descriptions which are robust to changing the bond dimension.

\bibitem{AKLT} I. Affleck et al., Phys. Rev. Lett. \textit{59}, 799 (1987).
\bibitem{Haldane83} F.D.M. Haldane, Phys. Rev. Lett. \textbf{50}, 1153 (1983), Phys. Lett. 93, \textbf{464} (1983).

\bibitem{White92} S.R.White, Phys. Rev. Lett. \textbf{69}, 2863 (1992).
\bibitem{Vidal03} G. Vidal, Phys. Rev. Lett. \textbf{91}, 147902 (2003); Phys. Rev. Lett. \textbf{98}, 070201 (2007).


\bibitem{Boyle78} L. L. Boyle and Kerie F. Green, Mathematical and Physical Sciences A \textbf{288}, 1351, pp. 237-269 (1978).
%

\bibitem{BLBQ} A. L\"auchli, G. Schmid, and S. Trebst, Phys. Rev. B \textbf{74}, 144426 (2006); Z.-X. Liu et. al., Phys. Rev. B \textbf{85}, 195144 (2012).




\bibitem{Gu09} Z.-C. Gu and X.-G. Wen, Phys. Rev. \textbf{B} 80, 155131 (2009).
\bibitem{Hu11} S. Hu, B. Normand, X. Wang, and L. Yu, Phys. Rev. \textbf{B} 84, 220402 (2011).
\end{thebibliography}
\end{document}